\documentclass[aps,pre,preprint,superscriptaddress]{revtex4-1}
\usepackage{graphicx}
\usepackage{amsmath}
\usepackage{amssymb}
\usepackage{psfrag}
\usepackage{natbib}

\usepackage{color}

\newcommand {\e} {\varepsilon}

\def \const{\mbox{const}}
\def \w {\omega}

\def \e {\varepsilon}
\def \vp {\varphi}

\def \ii {\text{i}}

\newcommand{\be}{\begin{equation}}
\newcommand{\ee}{\end{equation}}

\newcommand{\bea}{\begin{eqnarray}}
\newcommand{\eea}{\end{eqnarray}}

\begin{document}
	\title{Hub induced remote synchronization and desynchronization in complex networks of Kuramoto oscillators}
	
	\author{Vladimir Vlasov}
	\author{Angelo Bifone}
	\affiliation{Center for Neuroscience and Cognitive Systems,
Istituto Italiano di Tecnologia, Corso Bettini, 31, I-38068 Rovereto, Italy}
	\date{\today}
	
	\begin{abstract}

The concept of ``remote synchronization'' (RS) was introduced in [Phys. Rev. E 85, 026208 (2012)], where synchronization in a star network of Stuart-Landau oscillators was investigated. In the RS regime therein described, the central hub served as a transmitter of information between peripheral nodes, while maintaining independent dynamics that were asynchronous with the rest of the network. One of the key conclusions of that paper was that RS cannot occur in pure phase-oscillator networks. Here, we show that the RS regime can exist in networks of Kuramoto oscillators, and that hub nodes can actively drive remote synchronization even in the presence of a repulsive mean field. We apply this model to study the synchronization dynamics in complex networks endowed with hub-nodes, an ubiquitous feature of many natural networks. We show that a change in the natural frequency of a hub can alone reshape synchronization patterns, and switch from direct to remote synchronization, or to hub-driven desynchronization. We discuss the potential role of this phenomenon in real-world networks, including the Karate-club and brain connectivity networks.

	\end{abstract}

\maketitle

	

\section{Introduction}

	Synchronization of oscillatory units is a pervasive phenomenon that is responsible for the emergence of collective behaviors in natural and artificial systems~\cite{Pikovsky-Rosenblum-Kurths2001}. Entrainment of these dynamical systems depends on the characteristics of the individual oscillators, and on the nature and topology of the couplings that describe the interactions between oscillators. Perhaps the simplest representation of this phenomenon is the Kuramoto model~\cite{Kuramoto1975,Kuramoto1984}, where individual units are described as pure phase-oscillators interacting through phase-dependent couplings and characterized by a natural frequency~\cite{Acebron2005,Pikovsky2015}. In recent years, substantial emphasis has been put on the effects of the structure of the interaction network on synchronization~\cite{Arenas2008}. Indeed, the interplay between the dynamical and structural properties of complex networks of oscillators can generate interesting phenomena, including explosive synchronization~\cite{Gomez-Gardenes2011,Vlasov2015} and the emergence of cluster synchronization~\cite{Belykh2008,Pecora2014}.
	
	Recent findings by~\cite{Bergner2012} revealed an unexpected behavior dubbed ``remote synchronization" (RS) in star-like networks of oscillators, whereby unconnected peripheral oscillators can synchronize through a hub that maintains free, independent dynamics. The occurrence of RS is intriguing and somewhat counterintuitive, as it implies synchronization of oscillators that are not directly connected by structural links, nor by chains of entrained oscillators. Hence, RS seems to entail a ``hidden'' transfer of information between remote nodes in the network.
	
	Remote synchronization has been demonstrated experimentally in simple models, like star~\cite{Bergner2012} or ring~\cite{Minati2015} networks of oscillating electronic circuits, for example. Whether this phenomenon plays a relevant role also in complex networks, characterized by features like scale-freeness, small worldness, and the presence of rich clubs of highly connected nodes, remains the subject of active investigation~\cite{Gambuzza2013}. Indeed, the pervasive presence of hubs in real-world networks, i.e. star-like motifs embedded in otherwise sparsely connected networks, suggests that RS may significantly contribute to the synchronization patterns often observed in these systems. 

	Here, we apply a model of coupled phase-oscillators to investigate remote synchronization in the presence of hubs in complex networks. In the original paper on RS~\cite{Bergner2012}, it was suggested that remote synchronization requires the inclusion of free amplitudes in the model, and cannot be generated by simple Kuramoto oscillators. Using the analytical results from~\cite{Vlasov2015b}, we show that RS can also be observed in star-networks of pure phase-oscillators. Moreover, we implement a Kuramoto model to study the conditions for remote synchronization in real world networks while minimizing the need for parameterization of the oscillating units. We show that a degree-dependent distribution of natural frequencies, with hubs presenting a frequency off-set, can induce complex patterns of remote synchronization, and that shifts in a single hub's frequency can result in dramatic changes in synchronization patterns. Importantly, we demonstrate that the hub plays an active role in remote synchronization, rather than merely transferring information between peripheral nodes.

	To explore the potential relevance of hub-driven remote synchronization in real networks, we have chosen the Karate-club network~\cite{Zachary1977}, a prototypical social network endowed with a modular structure, and a brain connectivity network with structural hubs. Specifically, we leverage recent electrophysiological and structural connectivity data to model the dynamics of spontaneous activity in the macaque brain, and investigate the role of hubs in shaping patterns of synchronization. 

\section{Star network}

	We adopt the Kuramoto-Sakaguchi model~\cite{Sakaguchi1986} of star-like network of identical phase oscillators described in~\cite{Vlasov2015b}
	\be
		\label{gen.0}
		\begin{split}
			\dot{\vp_k}&=\omega+A\sin(\phi-\vp_k-\alpha),\quad
k=1\ldots N,\\
			\dot{\phi}&=\omega_0+{1\over N}\sum_{j=1}^N B
\sin(\vp_j-\beta-\phi).
		\end{split}
	\ee
	where $\phi$ denotes the phase of the hub (or leader) and $\vp_k$ the phases of the leaf oscillators. In this case, a synchronous solution will include constant phase differences between oscillators.  
	
	As shown in~\cite{Vlasov2015b}, the dynamics of this system is quite rich, and includes hysteretic transitions between asynchronous and synchronous states. Depending on the model's parameters, the system~(\ref{gen.0}) can have three different stable solutions.

	(i) In the region of relatively small absolute values of frequency mismatch $|\w-\w_0|$, the system~(\ref{gen.0}) has one stable synchronous solution that is stationary in the rotating reference frame. For this solution, the phases $\vp_k=\Phi$, $k=1\ldots N$ are identical, while the phase of the leader $\phi=\Phi-\Delta\Phi$, where $\Delta\Phi=\const$. Hence, the synchronous solution has non-zero but constant phase shift between the leader and the leaves. 

(ii) With increasing frequency mismatch, an asynchronous regime emerges,. Stability of the asynchronous solution depends on the sign of the expression $\text{sign}(\sin(\alpha+\beta))(\w-\w_0)$. For positive values, the asynchronous solution is stable, while for negative values it becomes unstable.

(iii)  When the frequency mismatch is too large to lock the phases of the leader and the leaves, the synchronous solution (i) turns into the RS regime, whereby all the leaf oscillators have same frequency and phase, while the leader's frequency and phase are different. In~\cite{Vlasov2015b} this is called ``synchronous limit cycle" solution. A specific condition for the RS regime to be stable is:
	\be
		\label{RS-in-star.cond}
		\text{sign}(\sin(\alpha+\beta))(\w-\w_0)<-\sqrt{A^2+B^2+2AB\cos(\alpha+\beta)\,}.
	\ee
	
\begin{figure}[tbh]	
		\centering
		\includegraphics[width=0.95\columnwidth, clip]{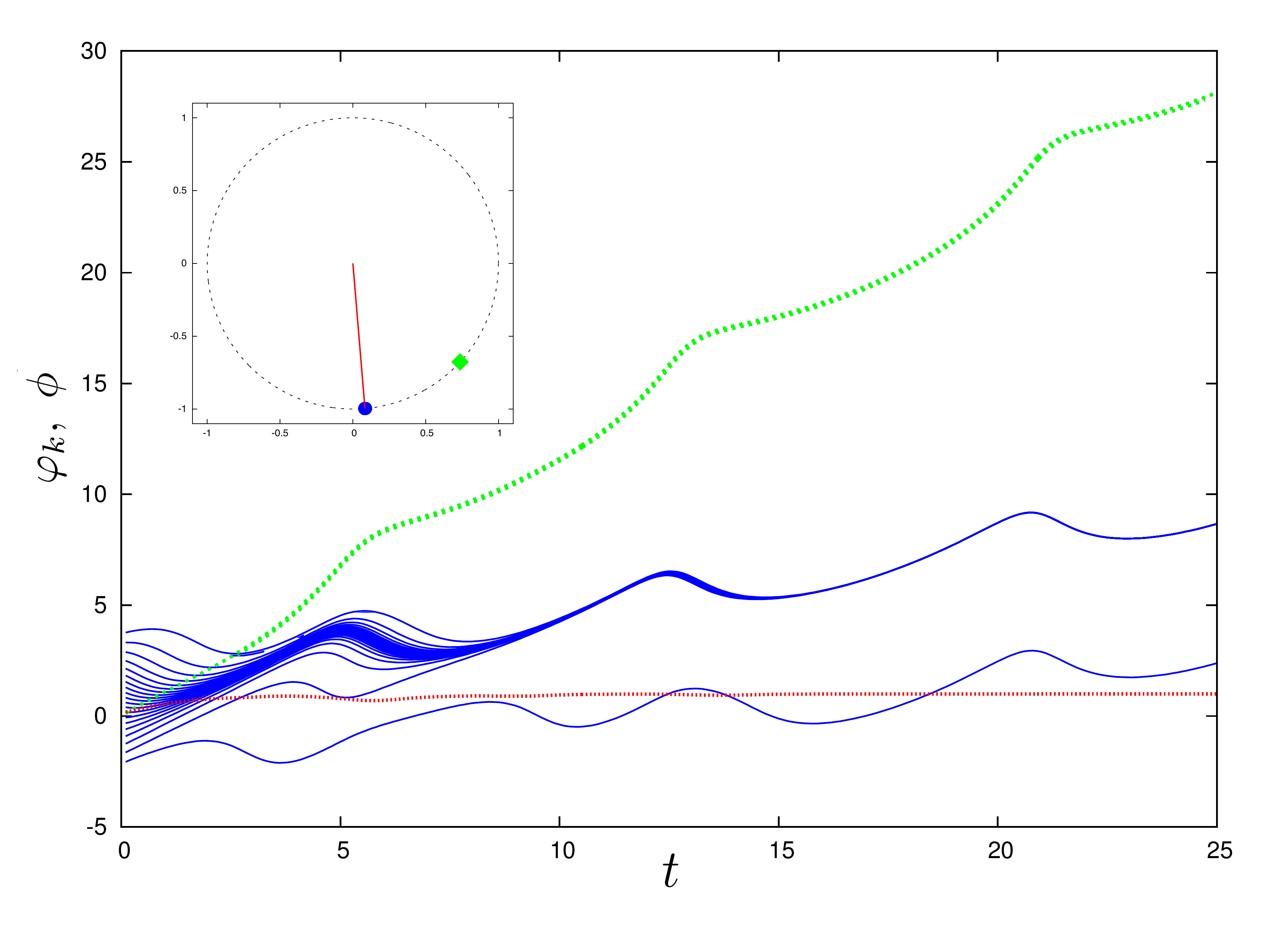}
		\caption{The simulation of the system~(\ref{gen.0}) for $N=20$ oscillators with the parameters $A=B=1$, $\alpha+\beta=0.6\pi$, $\w=0$ and $\w_0=1.4$. Time-dependence of the phase of the leader $\phi(t)$ (dashed green line), the phases $\vp_k(t)$ of the leaves  (solid blue lines) and their order parameter (dotted red line). Note that the difference between the blue branches is exactly $2\pi$. (Inset) Snapshot of the oscillators. The leaves are blue circles, the leader is green diamond, red line is the order parameter of the leaves.}
		\label{fig.RS-in-star}
	\end{figure}

	Fig.~\ref{fig.RS-in-star} demonstrate the RS solution of the system~(\ref{gen.0}) for the model parameters satisfying expression~(\ref{RS-in-star.cond}). This figure shows that, under these conditions, the difference between the synchronized leaves and the hub's phases builds up in time.

	From the brief description reported above and expression~(\ref{RS-in-star.cond}) it follows that the case of zero overall phase shift ($\alpha=-\beta$) is special. Indeed, stability analysis~\cite{Vlasov2015b} shows that in this case the asynchronous fixed point is neutrally stable. This implies that the RS regime is neutrally stable as well, thus making it difficult, if not impossible, to detect it numerically in a pure Kuramoto model with zero phase shift in the coupling term. This is the reason why the RS regime was not observed with pure Kuramoto oscillators in~\cite{Bergner2012}, where the authors concluded that RS can only be observed in the presence of an additional degree of freedom, like amplitude in the  Stuart-Landau equations. This additional degree of freedom was thought to be necessary for the appearance of RS regime, as it enables a hidden transfer of information through the amplitudes of the oscillators. However, the appearance of RS regime can arise from direct action of the hub on the leaves, as discussed below.

\subsection{Additional mean-field coupling}
	
	In order to show that a leader (or hub) can have a direct synchronizing or desynchronizing effect on peripheral nodes depending on the model's parameters, we consider a star network where the leaf oscillators are additionally subjected to a Kuramoto-Sakaguchi mean field. This system was partly studied in~\cite{Vlasov2015b} in the form:
	\be
		\label{m-f.gen.0}
		\begin{split}
			\dot{\vp_k}&=\omega+A\sin(\phi-\vp_k-\alpha)+{1\over N}\sum_{j=1}^NC\sin(\vp_j-\vp_k-\gamma),\quad
k=1\ldots N,\\
			\dot{\phi}&=\omega_0+{1\over N}\sum_{j=1}^N B
\sin(\vp_j-\beta-\phi).
		\end{split}
	\ee
	A complete analysis of this system~(\ref{m-f.gen.0}) for the entire range of parameters is beyond the scope of this paper. Here we will concentrate on particular solutions for the case of relatively weak mean field coupling. 

\begin{figure}[tbh]	
		\centering
		\includegraphics[width=0.95\columnwidth, clip]{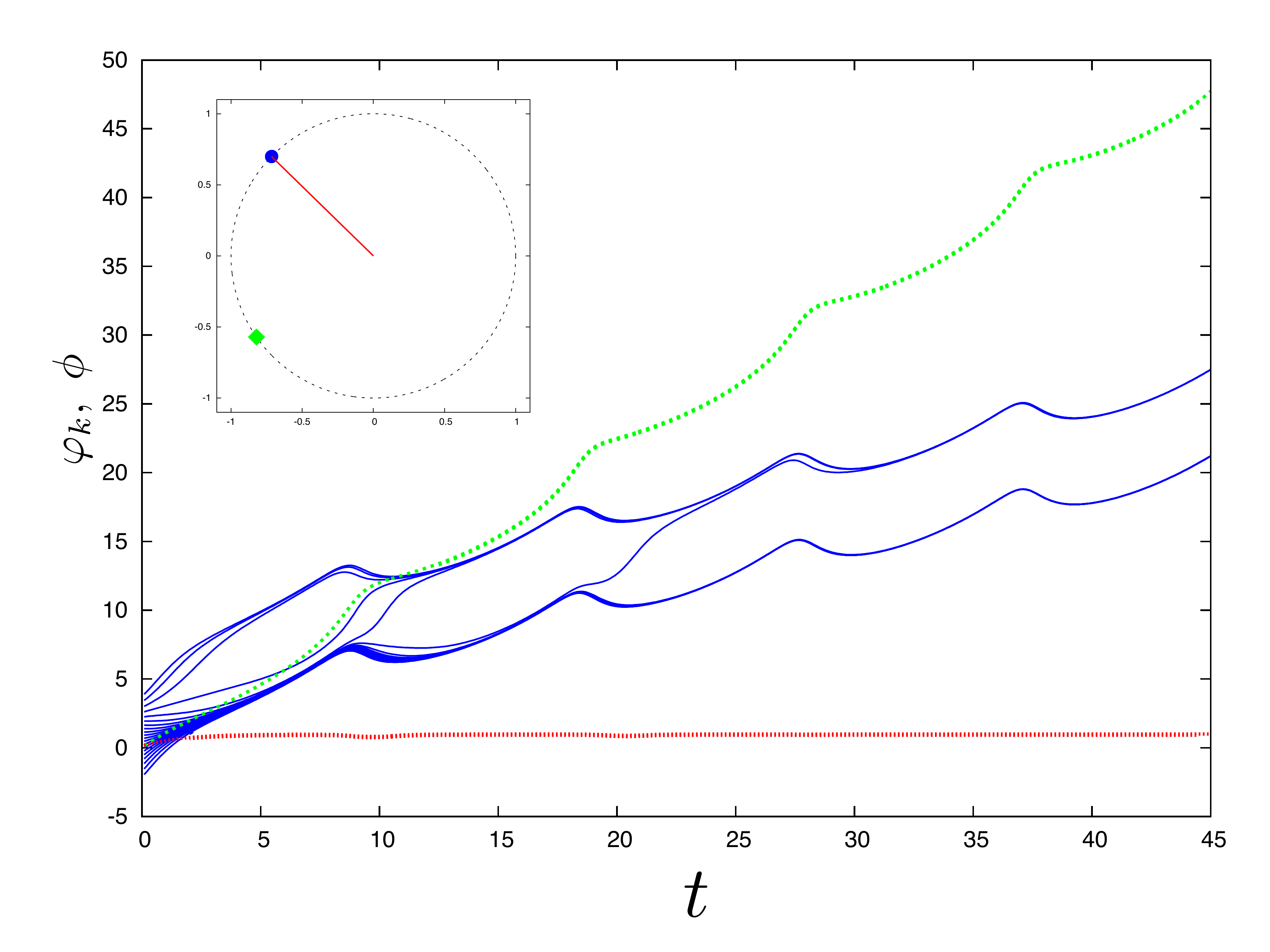}
		\caption{The simulation of the system~(\ref{m-f.gen.0}) for $N=20$ oscillators with the parameters $A=B=C=1$, $\alpha+\beta=\gamma=0.6\pi$, $\w=1$ and $\w_0=1.4$. Time-dependence of the phase of the leader $\phi(t)$ (dashed green line), the phases $\vp_k(t)$ of the leaves  (solid blue lines) and their order parameter (dotted red line). Note that the difference between the blue branches is exactly $2\pi$. (Inset) Snapshot of the oscillators. The leaves are blue circles, the leader is green diamond, red line is the order parameter of the leaves.}
		\label{fig.RS-in-star-Kur-rep}
	\end{figure}

(i) The asynchronous solution in the presence of attractive ($\cos\gamma>0$) mean field. In this regime the leader desynchronizes leaf oscillators in the bounded manifold of the positive frequency mismatch $\w-\w_0$ (see Fig.~\ref{fig.RS-in-star-Kur-rep}).
	. 
	
	\begin{figure}[tbh]	
		\centering
		\includegraphics[width=0.95\columnwidth, clip]{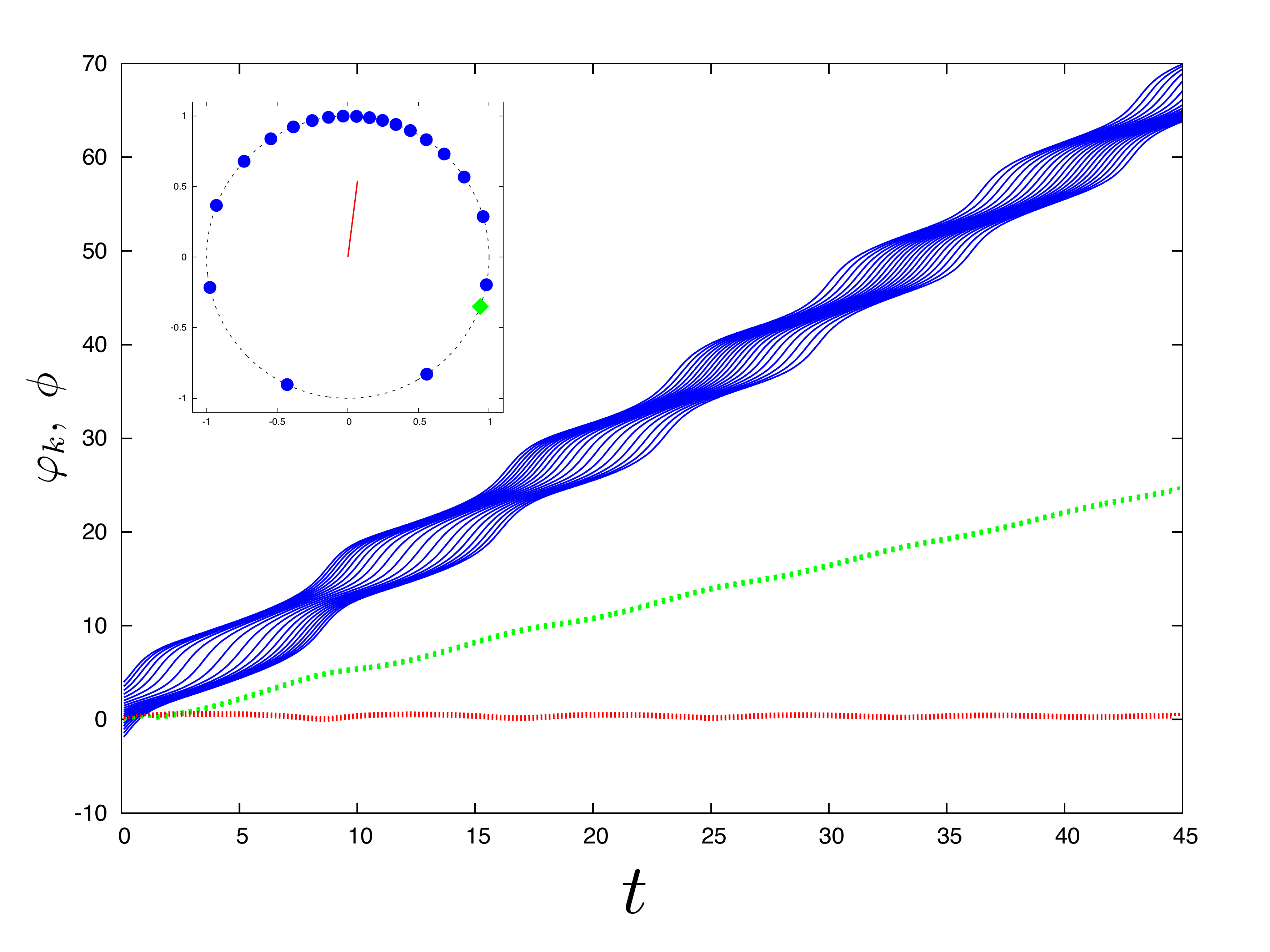}
		\caption{The simulation of the system~(\ref{m-f.gen.0}) for $N=20$ oscillators with the parameters $A=B=1$, $C=0.5$, $\alpha+\beta=0.6\pi$, $\gamma=0.2\pi$, $\w=2$ and $\w_0=0.6$. Time-dependence of the phase of the leader $\phi(t)$ (dashed green line), the phases $\vp_k(t)$ of the leaves  (solid blue lines) and their order parameter (dotted red line). (Inset) Snapshot of the oscillators. The leaves are blue circles, the leader is green diamond, red line is the order parameter of the leaves.}
		\label{fig.RS-in-star-Kur-attr}
	\end{figure}

	(ii) The synchronous solution in the presence of repulsive ($\cos\gamma<0$) mean field. In this regime, the leader synchronizes the leaves in the bounded region of the negative frequency mismatch $\w-\w_0$. As in the RS regime, the leader does not necessarily need to be synchronous with the leaves (see Fig.~\ref{fig.RS-in-star-Kur-rep}).
	
The above examples show that the leader can have a synchronizing action even in the presence of a desynchronizing mean field or, vice versa, a desynchronizing one in the presence of a synchronizing mean field. This is inconsistent with the notion that the hub serves the sole purpose of enabling information transfer between the peripheral nodes, and suggests that it directly drives the dynamics of the leaves.

\section{Karate-club network}

	In this section we investigate remote synchrony between oscillators in real world networks with hubs. As an example, we examine the Karate Club network~\cite{Zachary1977}, a widely studied social network in which certain nodes, corresponding to individuals with leading roles in a sports club, present a larger number of connections to other members. On this network we implement a Kuramoto-Sakaguchi model:
	\be
		\label{RS-in-net.sys}
		\dot{\vp_i}=\omega_i+{\e\over k_i}\sum_{j=1}^N A_{ij}\sin(\vp_j-\vp_i-\delta),\quad i=1\ldots N,
	\ee
	where $k_i=\sum_{j=1}^N A_{ij}$ is degree of the node $i$.

As shown in the previous section, if the sine of the cumulative phase shift is positive, the frequencies of the hubs must be sufficiently larger then the frequencies of the peripheral oscillators (leaves) in order to observe the RS regime. Therefore, we adopt the distribution of frequencies from~\cite{Gomez-Gardenes2011}, where the frequency of the oscillator is proportional to the degree  of the node $\w_i = k_i$ (by rescaling time the coefficient of proportionality can be absorbed in the coupling strength $\e$). As a measure of synchrony between nodes, we adopt the time-averaged order parameter called synchronization index:
	\be
		\label{RS-in-net.corr}
		r_{ij}=|\langle e^{\ii[\vp_i(t)-\vp_j(t)]}\rangle_t|,
	\ee
	where $\langle \cdot \rangle_t$ is an average over large period of time. Note that this order parameter is not sensitive to constant phase shifts, and takes large values when the average frequencies of two oscillators are similar. We consider nodes $i$ and $j$ to be synchronized if $r_{ij} > 0.75$.
	
	\begin{figure}[tbh]	
		\centering
		\includegraphics[width=0.95\columnwidth, clip]{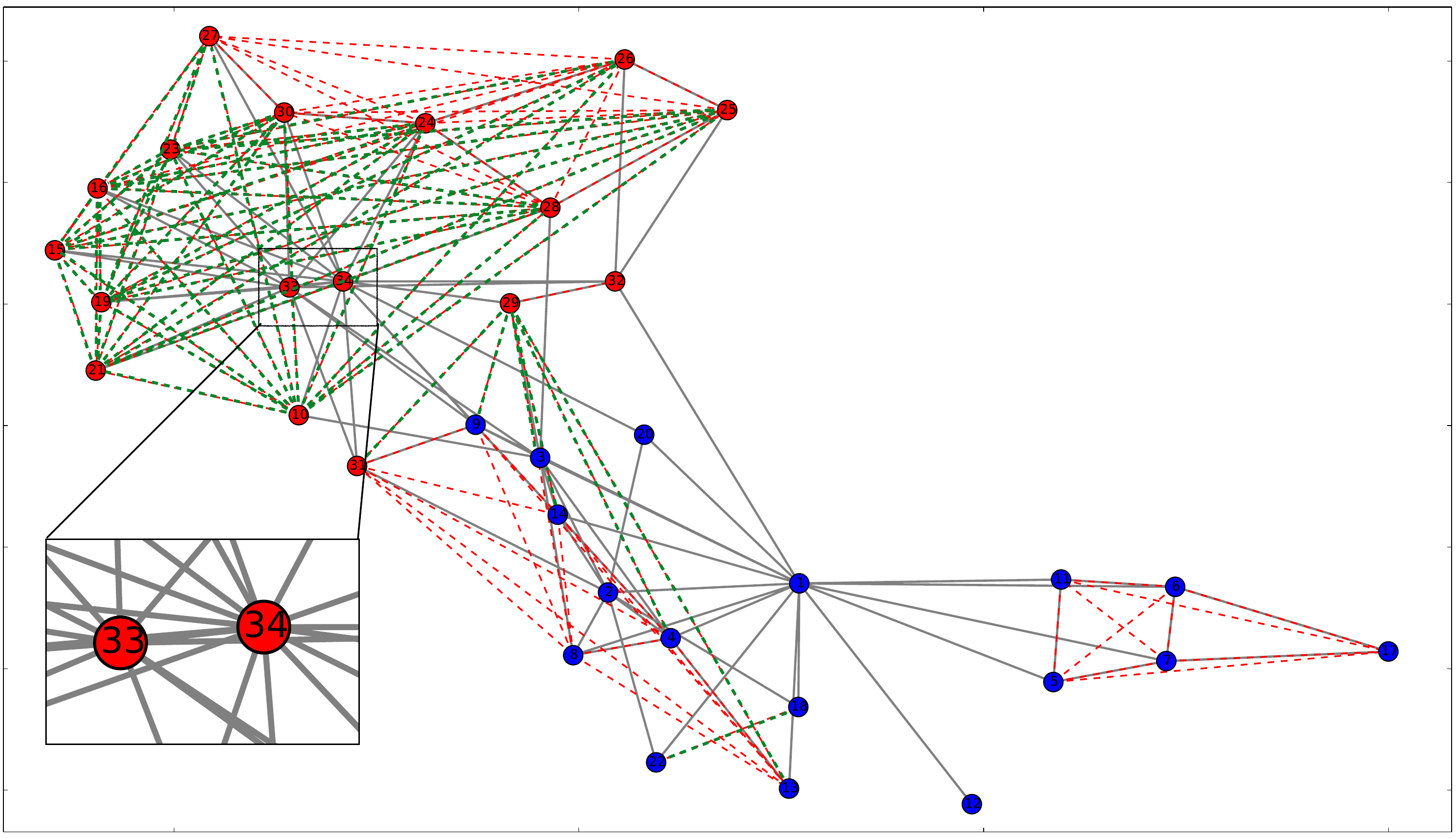}
		\caption{The simulation of the system~(\ref{RS-in-net.sys}) with the parameters $\e=5$, $\delta=0.2\pi$. Karate club network is shown by solid grey connections. Remote and direct synchrony links with $r_{ij}>0.75$ are shown by dashed green and red lines respectively.}
		\label{fig.Kar-club-RS}
	\end{figure}

	Figure~\ref{fig.Kar-club-RS} shows the results of the simulations on the Karate-club network, where grey lines denote structural links, and red and green dashed lines connect nodes that are directly or remotely synchronized, respectively. 

In the case of non-zero phase shifts $\delta$, we note that the highly connected hubs $33$ and $34$ generate a cluster of remotely synchronized nodes while remaining asynchronous with their leaf oscillators.
	
\begin{figure}[tbh]	
		\centering
		\includegraphics[width=0.95\columnwidth, clip]{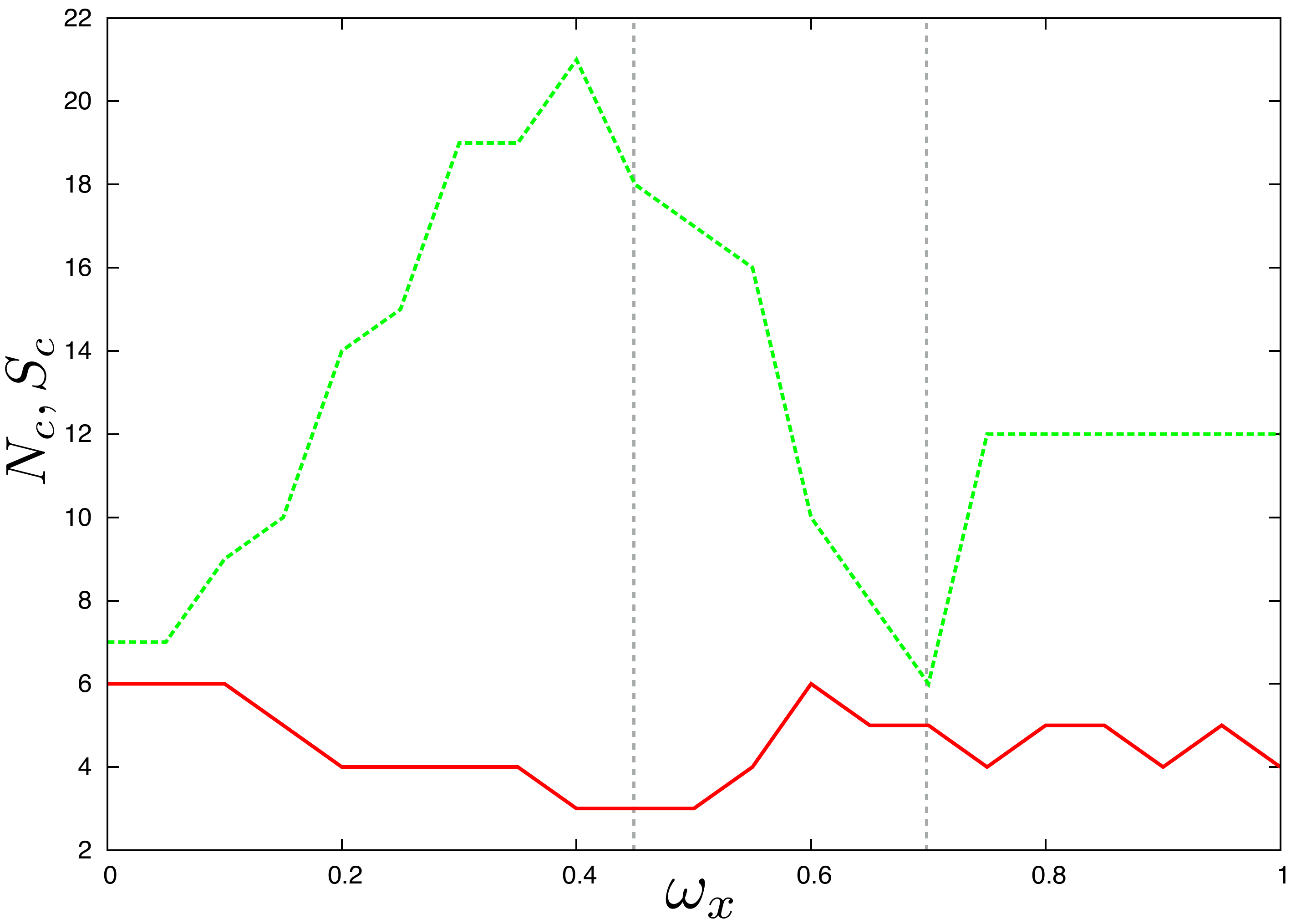}
		\caption{The dependance of the number of synchronized clusters $N_c$ (solid red line) and the size of the largest cluster $S_c$ (dashed green line) on the hub frequency multiplier $\w_x$.}
		\label{fig.Clust-Wx}
	\end{figure}

\begin{figure}[tbh]	
		\centering
		\includegraphics[width=0.95\columnwidth, clip]{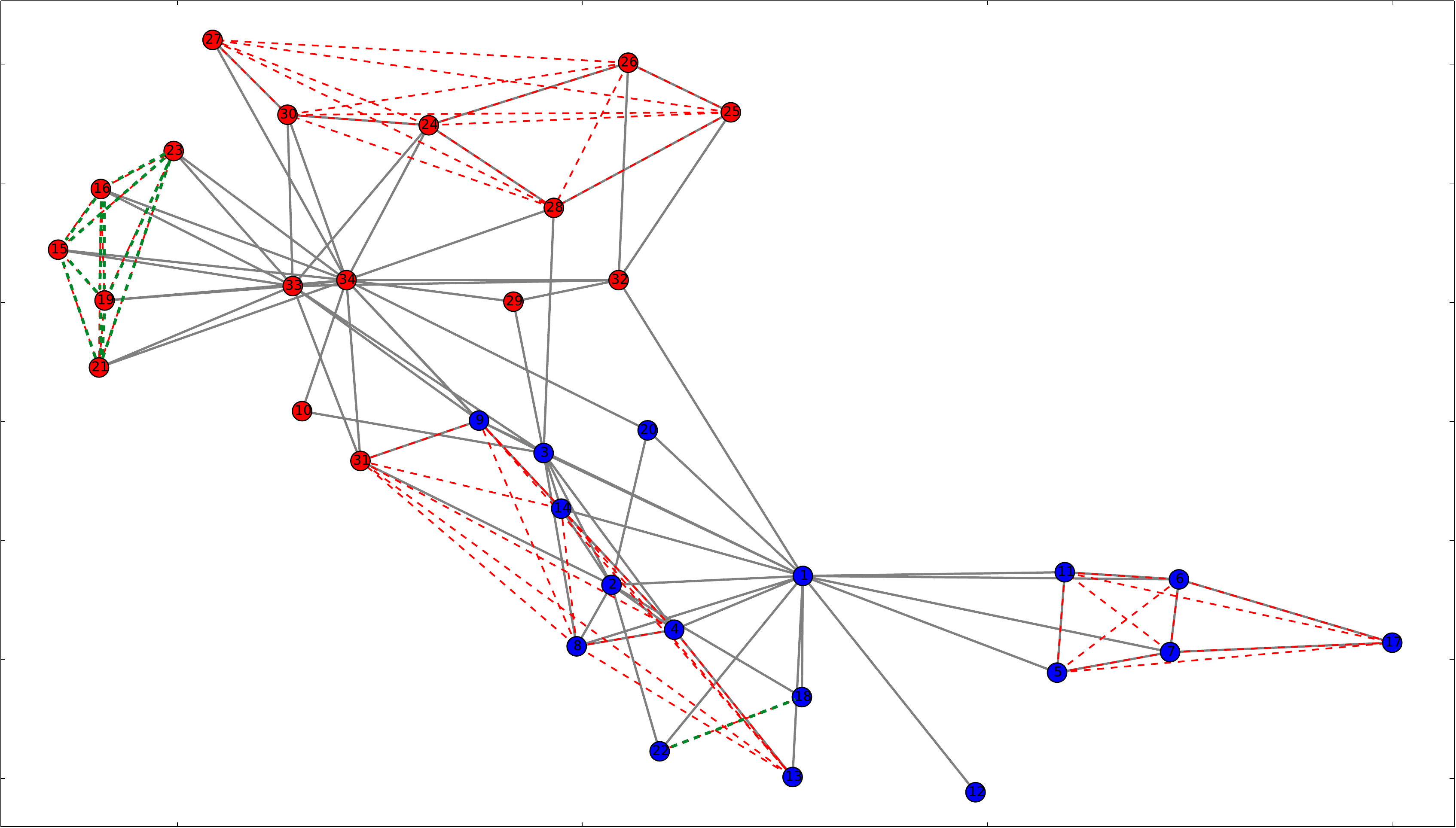}
		\caption{The same as Fig.~\ref{fig.Kar-club-RS} but for $\w_x=0.7$.}
		\label{fig.net-kc-wx-07}
	\end{figure}

	In order to further study how the frequency of hubs affects synchronization of the network, a multiplier $\w_x$ was introduced to selectively vary the frequency of the two nodes ($33$ and $34$). Fig.~\ref{fig.Clust-Wx} shows the  number of synchronized clusters of nodes and the size of the largest cluster as a function of the multiplier $\w_x$. For small values of $\w_x$, the number of clusters is quite large, and the size of the largest cluster is small. With increasing $\w_x$ the size of the largest cluster follows a bell shaped curve, with a maximum corresponding to $\w_x=0.7$. Conversely, the number of clusters shows a complementary behavior, albeit less pronounced.  
At $\w_x=0.7$ remote synchronization starts to emerge, comprising small basins of nodes (Fig.~\ref{fig.net-kc-wx-07}). Above $\w_x=0.7$, the size of the largest cluster increases sharply and remains constant for increasing values of $\w_x$. In this regime, the frequencies of the hubs are large enough to remotely synchronize their leaf nodes (see Fig.~\ref{fig.Kar-club-RS}). 
		

\begin{figure}[tbh]	
		\centering
		\includegraphics[width=0.95\columnwidth, clip]{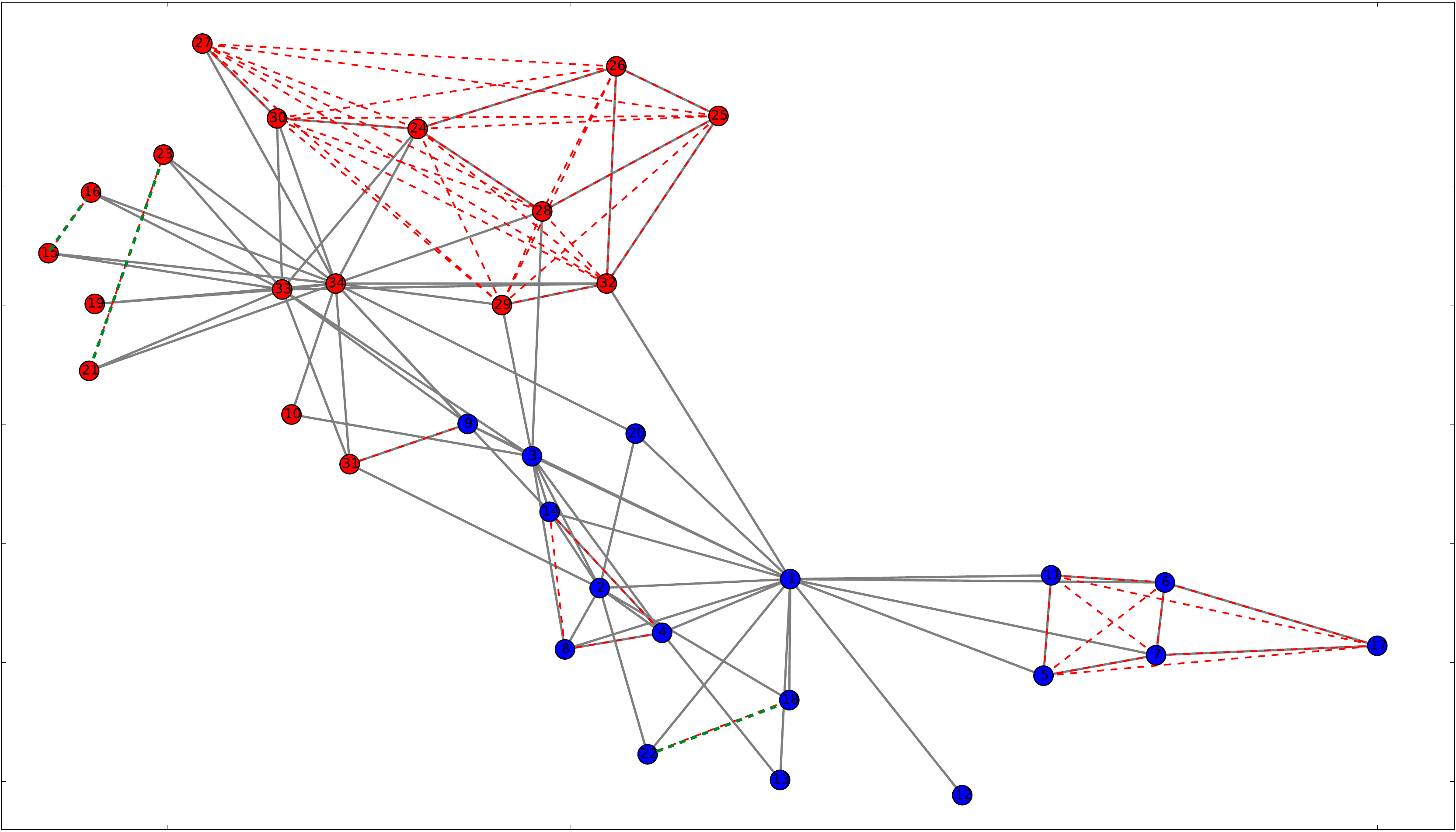}
		\caption{The same as Fig.~\ref{fig.Kar-club-RS} but for $\delta=0$.}
		\label{fig.net-kc-delt-0}
	\end{figure}

	In the case of zero phase shift $\delta=0$, occasional pairwise correlations may appear (e.g. between nodes $15$ and $16$ in Fig.~\ref{fig.net-kc-delt-0}) depending on the initial conditions, but extended clusters of remotely synchronized oscillators do not emerge.

\section{Application to Macaque brain}

	The interplay between structural connections and synchronization of dynamical processes on networks is reminiscent of the concept of functional connectivity in the realm of neuroscience. In this context, functional connectivity is defined in terms of correlations or coherence between oscillatory behaviors observed, e.g., in the electrical or hemodynamic spontaneous activity of the brain. While harmonic oscillators are an oversimplification of the complex processes underlying neural or metabolic oscillatory activity, they are thought to capture the basic principles of the synchronization phenomena.

	The question we intend to address in the following is whether remote synchronization may play a role in determining patterns of correlated activity in brain networks as observed, e.g., in EEG or MEG neuroimaging experiments in the brain under resting conditions. The key ingredients for this phenomenon to emerge are:
(i)	the presence of hubs within brain networks of structural connectivity;
(ii)	a phase shift between remote nodes resulting from a delay in the interaction terms;
(iii)	a difference in the oscillatory frequency of hubs with respect to their peripheral nodes.

	A number of studies (see~\cite{VandenHeuvel2013,Sporns2013} for recent reviews) based on Diffusion Tensor Magnetic Resonance Imaging have shown that certain brain regions are connected by white matter tracts to many other brain areas, thus behaving as structural hubs. This has been observed in humans~\cite{VandenHeuvel2011} as well as in other species, including non-human primates and rodents~\cite{Sporns-Honey2007}. Ex-vivo studies using anterograde or retrograde tracers corroborate this evidence in experimental laboratory animals like the macaque. Hence, condition (i) appears to be fulfilled.

	Finite signal propagation speed in axons generates distance-dependent time delays in the interaction terms in brain networks ~\cite{Breakspear2010, Cabral2011, Cabral2014}. In a model of coupled phase oscillators, these delays can be represented as phase shifts proportional to internodal distance~\cite{Breakspear2010}, as in condition (ii).

	Finally, condition (iii) has been recently addressed in a meta-analysis of electrophysiology experiments in the macaque, showing anatomical dependence of spontaneous oscillations of populations of neurons in a number of brain areas as measured by invasive electrophysiology~\cite{Murray2014}. Comparison with degree distribution in macaque~\cite{Sporns-Honey2007} shows that the fast nodes from~\cite{Murray2014} are also structural hubs.

In the light of this evidence, we have modeled the synchronization phenomenon in the macaque brain using experimental data and empirically determined parameters from the literature.

For our simulations, we adopted the structural connectivity graph described in ~\cite{Markov2014,Ercsey-Ravasz2013}, where connectivity data was obtained by retrograde tracer injections in 29 areas of the macaque cerebral cortex. Extrinsic fraction of labeled neurons (the ratio between the number of labeled neurons in the source area over the total number of labeled cortical neurons extrinsic to the injected area) for each pathway determines the weight of the connection between areas. The 29 by 29  connectivity matrix was thresholded by percolation analysis of the giant component~\cite{Gallos2012} and binarized.

Simulations were performed based on this connectivity graph by adopting the model~(\ref{RS-in-net.sys}) without normalization of the coupling strengths by node degree. Following~\cite{Breakspear2010,Vlasov2014, Vlasov2015b} we assumed that the phase shifts in the coupling terms are proportional to the distances between nodes. Internodal distances were taken from~\cite{Ercsey-Ravasz2013} as the length of the shortest trajectory interconnecting areas via the white matter, approximating the axonal distance.

Murray et.al.~\cite{Murray2014} collected measurements of timescales of intrinsic fluctuations in spiking activity in different areas of the macaque brain. In~\cite{Chaudhuri2015}, the selection of regions was extended using modeling of the macaque neocortex. Specifically, Chaudhuri et al.~\cite{Chaudhuri2015} calculated autocorrelation functions of area activity in response to white-noise input to all areas. Time constants of the decay of autocorrelation were calculated for all nodes included in our model. The dominant time constants in various areas of the network were extracted by fitting single or double exponentials to the autocorrelation. In case of double exponentials, the timescales of the two components were calculated as the weighted average of the two time constants. For the frequencies in our model we took inverse timescales. 

\begin{figure}[tbh]	
		\centering
		\includegraphics[width=0.95\columnwidth, clip]{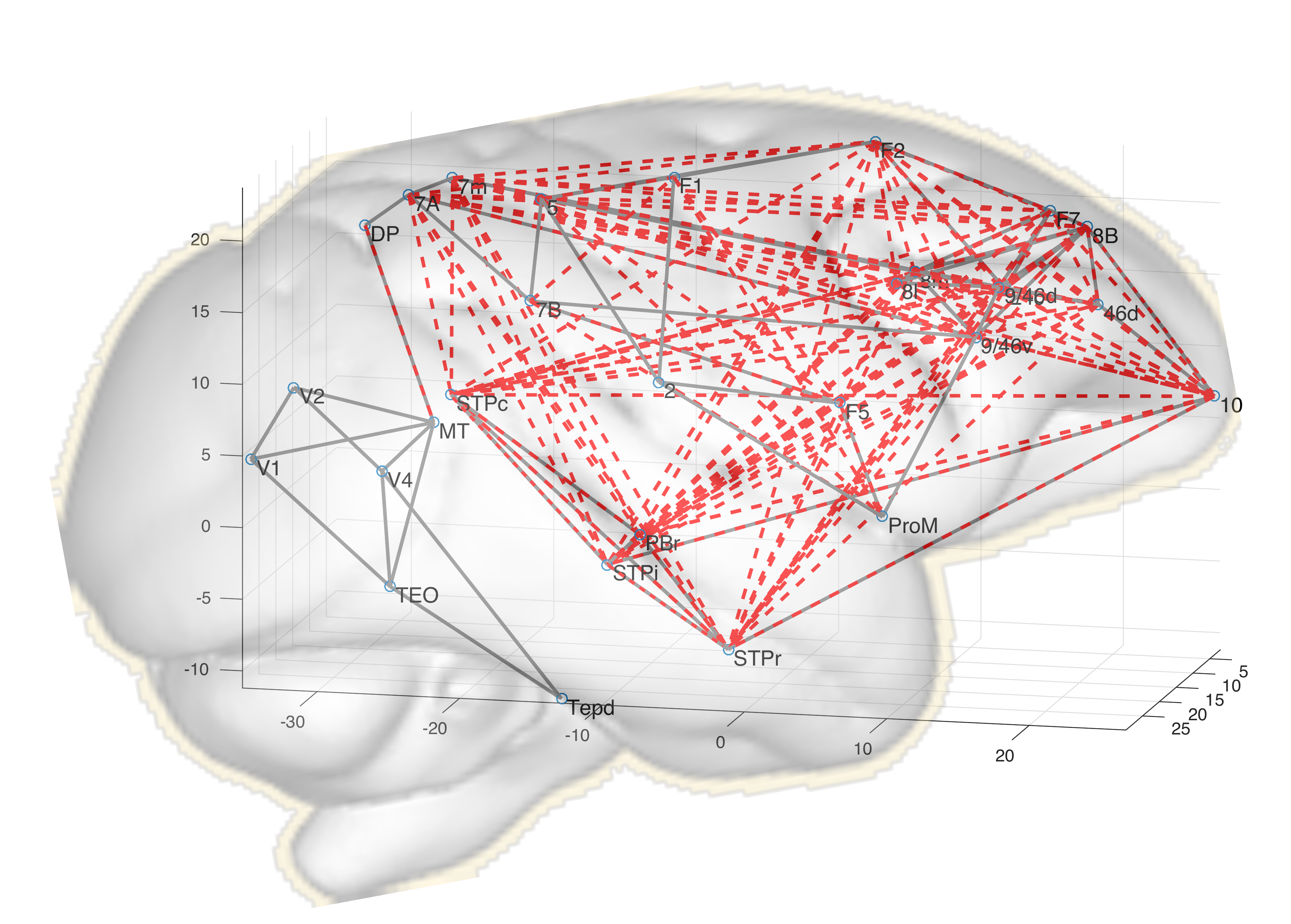}
		\caption{Simulation of the Macaque connectome. Structural links are shown with grey lines. Functional links with $r_{ij}>0.75$ are shown by dashed red lines. Names of the nodes are labeled on the figure. Frequencies are taken as inverse timescales for resting-state case from~\cite{Chaudhuri2015}, e.g. $\w_{10} = 1/3.17$ for the node 10. The other parameters are $\e=0.85$, $\delta_{ij} = 2\pi D_{ij}\nu_s/c$, where $\nu_s = 40$ Hz, $c=10$ m/s and $D_{ij}$ is the internodal distance matrix taken from~\cite{Ercsey-Ravasz2013}.}
		\label{fig.Macaque-norm}
	\end{figure}

\begin{figure}[tbh]	
		\centering
		\includegraphics[width=0.95\columnwidth, clip]{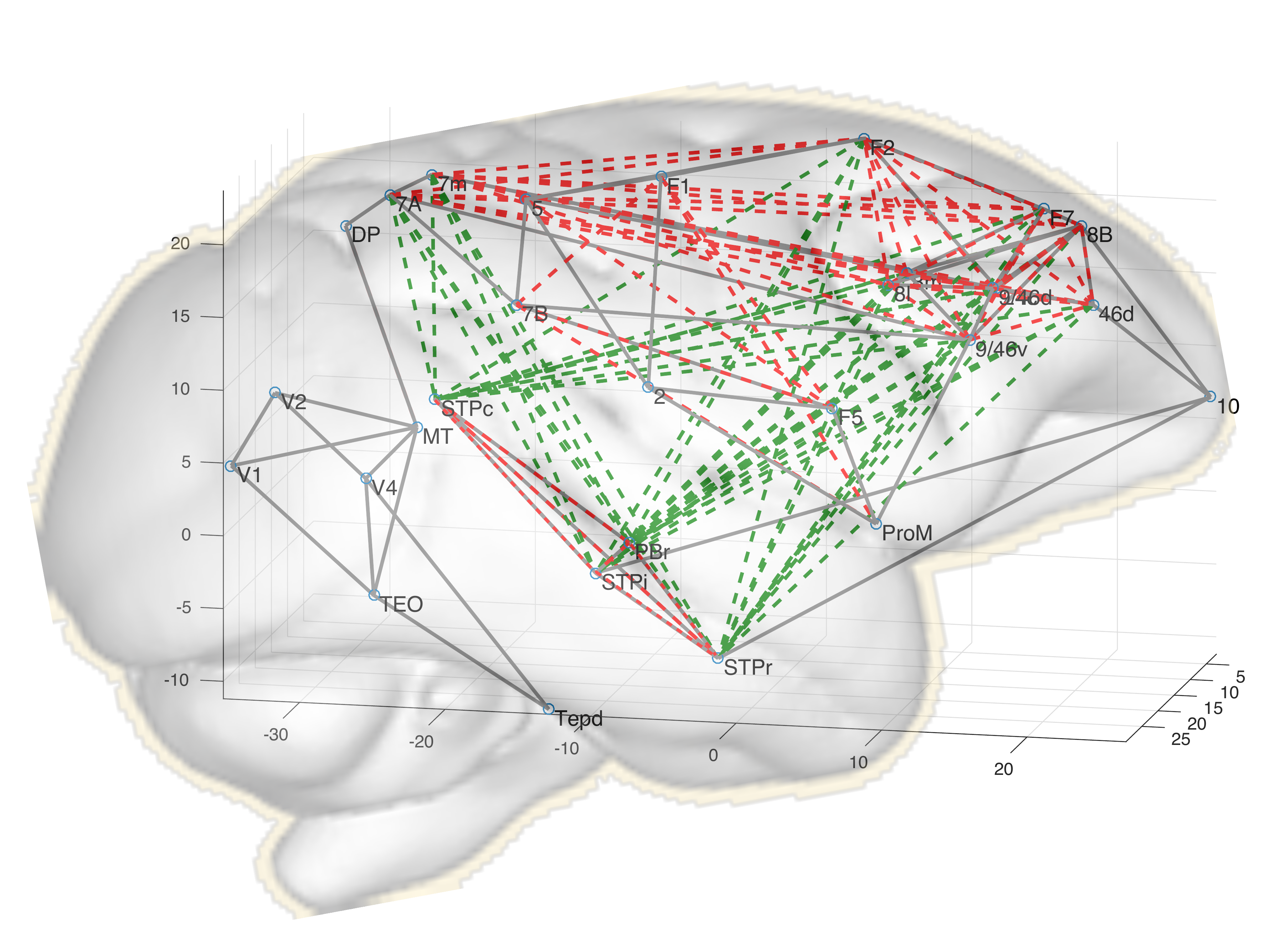}
		\caption{The same as Fig.~\ref{fig.Macaque-norm} but the frequency of the node 10 is larger $\w_{10} = 1/0.185$. Remotely synchronized functional links are shown by dashed green lines.}
		\label{fig.Macaque-rs}
	\end{figure}

\begin{figure}[tbh]	
		\centering
		\includegraphics[width=0.95\columnwidth, clip]{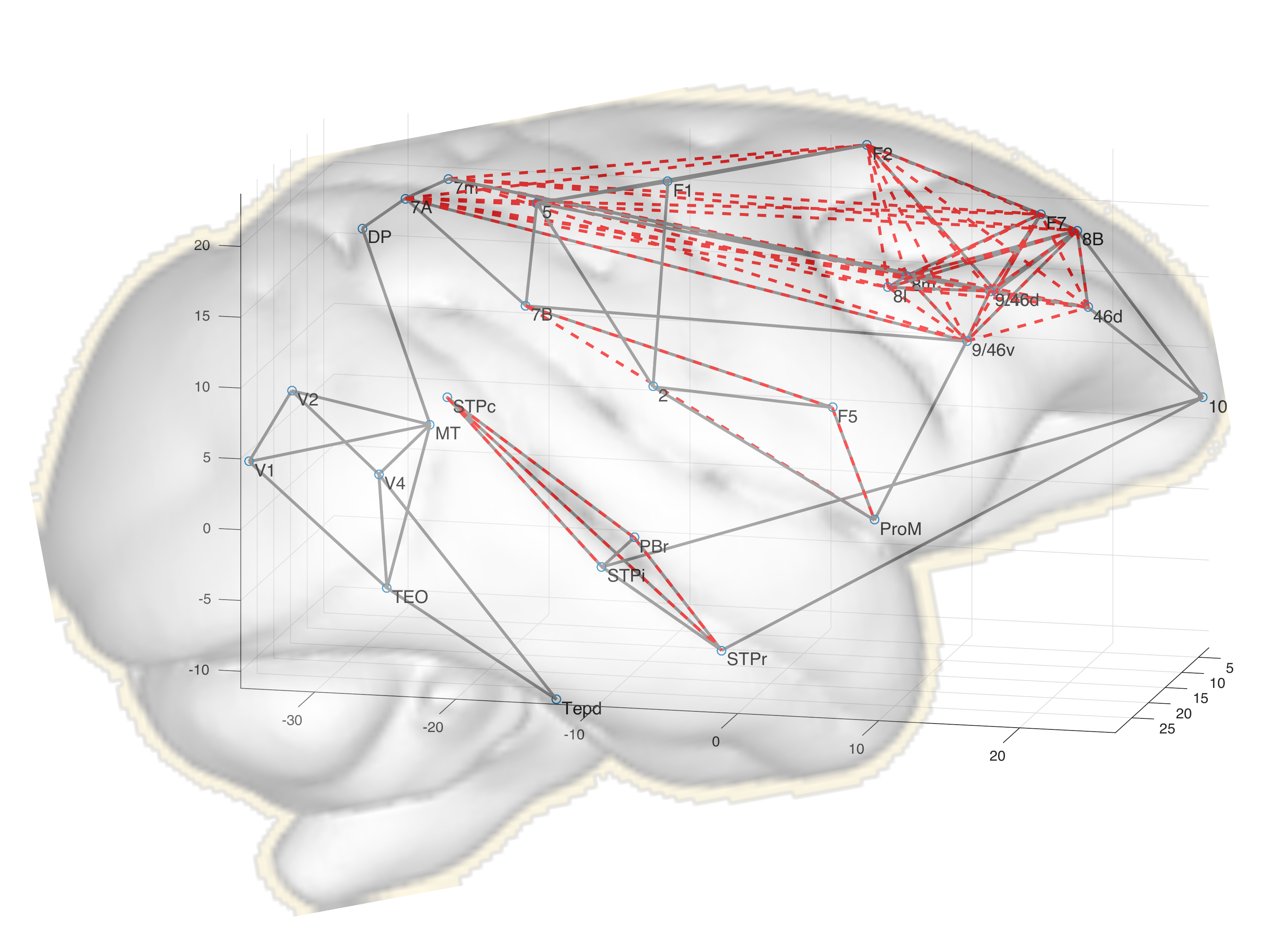}
		\caption{The same as Fig.~\ref{fig.Macaque-norm} but the frequency of the node 10 is even larger $\w_{10} = 1/0.0625$.}
		\label{fig.Macaque-asynch}
	\end{figure}

The results of our simulations are shown on Fig.~\ref{fig.Macaque-norm}. In the structural connectivity network, node in the prefrontal cortex denoted as $10$ plays the role of a structural hub that connects areas in the frontal and temporal cortices. Using the frequency distribution calculated from~\cite{Chaudhuri2015}, we obtain a wide pattern of synchronization (red dashed lines), including all frontal and temporal cortices.  When we switch the frequency of node 10 to that of the fast component associated with this node~\cite{Chaudhuri2015}, coherent synchronization among those areas is preserved, despite the fact that node 10 becomes asynchronous with the rest of the cluster, consistent with the definition of remote synchronization (green dashed lines in Fig.~\ref{fig.Macaque-rs}). When we further increase the frequency of the oscillator associated with node $10$, frontal and temporal areas become functionally disconnected, and form two separate basins of synchronized nodes. Hence, a switch between the regime of normal synchronization~(Fig.~\ref{fig.Macaque-norm}), remote synchronization~(Fig.~\ref{fig.Macaque-rs}) and asynchrony~(Fig.~\ref{fig.Macaque-asynch}) can be driven by a single parameter in the model, namely the hub's frequency.

This finding is particularly interesting as it may indicate a new and potentially important mechanism in the formation of functional connectivity patterns in the brain. Here, we show that clusters of synchronized areas can result from the activity of a structural hub region that does not appear to be functionally connected to these areas. This is consistent with the observation that structural and functional hubs in the brain do not necessarily coincide~\cite{Osada2015}. Importantly, changes in the frequency of spontaneous fluctuations in a hub region can dynamically reconfigure patterns of functional connectivity in the brain, switching between the regimes of direct and remote synchronization, or actively desynchronizing functional modules even when couplings between nodes remain attractive.
While the factors enabling this mechanism appear to be present in the brain, there is no direct evidence to support this phenomenon. However, this hypothesis can be experimentally tested. By way of example, it may be envisaged that optogenetic technology~\cite{Fenno2011}, whereby photosensitive proteins are expressed in specific populations of neurons, could provide a means to drive the oscillatory behavior of selected hub regions while measuring patterns of synchronized activity across the brain.


\section{Conclusions}
	
	In the first part of this paper we have summarized and emphasized the analytical results obtained in~\cite{Vlasov2015b} for the Kuramoto-Sakaguchi model on star networks. Specifically, we have explicitly shown that ``remote synchronization" (RS) can be observed in systems of Kuramoto phase oscillators if a non-zero phase shift is introduced in the coupling terms. For pure star networks, the RS regime is stable when a sufficiently large frequency mismatch is imposed between leaves and a hub. Under these circumstances, the hub can exert a synchronizing action even in the presence of an additional repulsive mean field acting on the peripheral nodes. Hence, the hub does not simply enable transfer of information between the leaves, but it actively drives synchronization while remaining asynchronous with the rest of the network. Conversely, the hub can actively desynchronize the leaves in the presence of an attractive mean field.

	In the second part we have explored the role of this mechanism in the synchronization of complex networks. As an example, we have chosen the Karate-club network, a prototypical social network in which a few highly connected individuals play the role of hubs. When a degree-dependent distribution of natural frequencies is introduced in the model, remote synchrony emerges and plays a substantial role in the formation of clusters of synchronized nodes within the network.

	Finally, we have explored the potential role of this mechanism in brain connectivity networks. Indeed, the prerequisites for hub-driven remote synchronization appear to exist in the brain, including the presence of structural hubs, delays in the coupling between nodes, and region-dependent frequency of spontaneous fluctuations. We have leveraged connectomic data from the macaque brain and recent electrophysiological measurements to derive a dynamical model of the population-level electrical activity in the macaque cortex. Our simulations show that a change in the intrinsic frequency of a hub can dramatically reshape the synchronization patterns, shifting from direct to remote synchronization, and to a hub-driven desynchronization regime. This experimentally testable hypothesis may explain the mismatch between structural and functional hubs sometimes observed in brain connectivity networks.

\section*{Acknowledgments}

We wish to thank Nikola T. Markov and Maria Ercsey-Ravasz for providing brain connectivity data. We also acknowledge interesting discussions with Ludovico Minati, Camillo Padoa-Schioppa, Cecile Bordier, Matteo Caffini, Carlo Nicolini, Michael Rosenblum, Arkady Pikovsky and Maxim Komarov.
This project has received funding from the European Union's Horizon 2020 research and innovation program under grant agreement No 668863. 

%

\end{document}